\documentclass[12pt]{iopart}

\begin{document}

\title[Exact Results for the Kepler Problem in General Relativity]{Exact Results for 
the Kepler Problem in General Relativity}

\author{K A Hall}

\address{65 Rosman Road, Thiells, NY 10984, USA}
\ead{ka\_hall@msn.com}
\begin{abstract}
Exact results are derived, specifically the perihelion shift and the Kepler orbit, for a bound test particle in the Schwarzschild metric with 
cosmological constant $\Lambda=0$. A series expansion, of $\Delta\phi=2(2(1-\frac{2M}{p}(3-e))^{-1/2}K(\frac{4eM/p}{1-\frac{2M}{p}(3-e)})-\pi)$, 
the exact perihelion shift, admits the standard approximation $\Delta\phi=\frac{6M\pi}{p}$ as the leading order term. In a similar 
fashion, a series expansion of the exact Kepler orbit, represented by a Jacobi elliptic function, gives $u(\phi)=\frac{1+e\cos\phi}{p}$ to first 
order. The results are valid for $M/p<\frac{1}{2(3+e)}$ or $r_s<\frac{p}{3+e}$.
\end{abstract}

%\submitto{\CQG}
\maketitle

\section{Introduction}
It is the intent of this paper to derive an exact representation for the perihelion shift $\Delta\phi$, and the orbital equation $r(\phi)$, 
for a bound test particle, of mass {\it m}, following a time-like geodesic in a Schwarzschild metric about a central mass {\it M}, 
where $m\ll M$. Einstein, in applying his new theory of General Relativity, addressed this problem, in 1915, to explain the perihelion shift 
of Mercury \cite{einstein}. For the specific case of Mercury's orbit around the Sun, where $M_\odot/p\sim 10^{-8}\ll1$, valid 
approximations, for $\Delta\phi$ and $r(\phi)$, are easily obtained using perturbation techniques. Therefore, an exact solution would be useful 
in situations where the expansion parameter $M/p$ is much larger, e.g. bound elliptical orbits 
around a massive black hole. In the above ratio, and henceforth, the semi-latus rectum $p$ has been scaled by the speed of light {\it c}, and 
the Sun's mass  $M_\odot$ has been scaled by $G/c^3$. We will use this standard convention, unless otherwise noted, to represent length and mass 
as seconds, specifically $r(cm)/c\rightarrow r(s)$ and $GM(g)/c^3\rightarrow M(s)$. 

Although an exact expression is known \cite{kraniotis}, e.g. Kraniotis and Whitehouse (2003), for the bound Kepler orbit and the perihelion shift, 
given as a function of the Weierstrass function 
${\it P}(z)={\it P}(z+2\omega_1)={\it P}(z+2\omega_2)$, and its half-periods $\omega_{1,2}$, respectively, it is the goal of this paper to 
present a compact representation using the eccentricity $e$ and the semi-latus rectum $p$. An exact solution, for the bound Kepler orbit, using the Jacobi elliptic function, was first presented in 1959 by C.G. Darwin \cite{darwin}.
\section{Calculations}

\subsection{Perihelion Shift}

For a test particle, of mass {\it m}, orbiting a central mass {\it M}, where $m\ll M$, in the space-time of the Schwarzschild metric, it is 
straightforward to show that $u(\phi)=1/r(\phi)$ satisfies
\begin{eqnarray}
\left(\frac{du}{d\phi}\right)^2=\frac{E^2}{L^2}-(1-2Mu)\left(\frac{m^2}{L^2}+u^2\right) \label{eq1}
\end{eqnarray}
where $\phi$ is the azimuthal angle, and $r(\phi)$ is the radial distance from the central mass M at the origin \cite{einstein,magnan}. The 
conserved quantities, the energy {\it E}, and the angular momentum {\it L}, are uniquely determined once the initial conditions, $u(\phi=0)$ 
and $\frac{{\rm d}u}{{\rm d}\phi}|_{\phi=0}$ are chosen. For bound orbits, it is more convenient to rewrite equation (\ref{eq1}) in 
terms of $u_p$ and $u_a$,
\numparts
\begin{eqnarray}
\fl\left(\frac{du}{d\phi}\right)^2=(u_p+u_a)(1-2Mu_p)(1-2Mu_a)/2M\nonumber\\-(1-2Mu)
\left[u_p+u_a-2M(u_p^2+u_a^2+u_pu_a)+2Mu^2\right]/2M{\rm ,}\\ 
\fl{\rm where}\qquad\frac{E^2}{L^2}-(1-2Mu_p)\left(\frac{m^2}{L^2}+u_p^2\right)=0\\
\fl{\rm and}\qquad\quad\frac{E^2}{L^2}-(1-2Mu_a)\left(\frac{m^2}{L^2}+u_a^2\right)=0\label{eq2}
\end{eqnarray}
\endnumparts
since the radial velocity must vanish at $u_p$ and $u_p$ \cite{magnan,book1}. This technique allows us to incorporate $E$ and $L$ into the endpoints of the range for $u$. The perihelion and aphelion, $r_p$ and $r_a$, 
and their inverses, satisfy
$$r_p\leq r\leq r_a\,{\rm ,}\; u_a\leq u\leq u_p\,\qquad\mbox{where}\qquad u_p=1/r_p\,{\rm ,}\; u_a=1/r_a\,{\rm .}$$
We now factor equation (2a), to obtain
\begin{eqnarray}
\left(\frac{du}{d\phi}\right)^2=2M(u-u_p)(u-u_a)(u-u_3)\,{\rm ,} \label{eq3}
\end{eqnarray}
and define
\begin{eqnarray}
u_3=\frac{1}{2M}-(u_p+u_a)\,{\rm .}\label{eq4}
\end{eqnarray}
As expected, equation (\ref{eq3}) equals zero when $u=u_p$ and $u=u_a$, since we expect the radial velocity to vanish at 
the perihelion and aphelion. These turning points correspond to the energy being equal to the effective potential energy.

In equation (\ref{eq3}), the product $(u-u_p)(u-u_a)\leq0\; \forall\; u\in[u_a,u_p]$, and, since we require a real radial velocity, this 
implies that $u-u_3<0\; \forall\; u\in[u_a,u_p]$. This condition implies an upper limit for the central mass $M$, represented by equation (\ref{eq5}):
\begin{eqnarray}
\fl \qquad\qquad\qquad\qquad\; u-u_3<0{\rm ,}\quad\mbox{for}\quad &M<\frac{1}{2(2u_p+u_a)}\label{eq5}\\
\fl-(u_p-u_a)<u-u_3<u_p-u_a{\rm ,}\quad\mbox{for}\quad\frac{1}{2(2u_p+u_a)}<&M<\frac{1}{2(2u_a+u_p)}\nonumber\\
\fl \qquad\qquad\qquad\qquad\; u-u_3>0{\rm ,}\quad\mbox{for}\quad &M>\frac{1}{2(2u_a+u_p)}\;{\rm .}\nonumber
\end{eqnarray}
The condition, in equation (\ref{eq5}), may also be viewed as a lower limit imposed upon the radial 
parameters, i.e. $r_s<1/(2u_p+u_a)$, where $r_s=2M$ is the Schwarzschild radius.

Expanding the product $(u-u_p)(u-u_a)$, and completing the square, gives a form that is now exactly integrable,
\begin{eqnarray}
\fl\left(\frac{du}{d\phi}\right)^2=2M\left((u-f)^2-b^2\right)
\left(u-f-\left(\frac{1}{2M}-3f\right)\right)=2M(y^2-b^2)(y-d){\rm ,} \label{eq6}
\end{eqnarray}
where  
\numparts
\begin{eqnarray}
f&=\frac{u_p+u_a}{2}\\
b&=\sqrt{f^2-u_pu_a}=\frac{u_p-u_a}{2}\\
y&=u-f\\
d&=\frac{1}{2M}-3f=\frac{1}{2M}-\frac{3}{2}(u_p+u_a){\rm .}\label{eq7} 
\end{eqnarray}
\endnumparts
Solving for ${\rm d}\phi$, and integrating both sides gives 
\begin{eqnarray}
\fl\phi+c_1=\frac{1}{\sqrt{2M}}\int\frac{{\rm d}y}{\sqrt{(y^2-b^2)(y-d)}}=
\frac{2}{\sqrt{2M(b+d)}}~F\left(\sin^{-1}\left(\sqrt{\frac{y+b}{2b}}~\right),\frac{2b}{b+d}\right)\label{eq8}
\end{eqnarray}
where $F(\varphi,m)$ is an elliptic integral of the first kind, defined by
\begin{eqnarray}
\fl F(\varphi,m)\equiv F(\varphi|m)=\int^\varphi_0{\rm d}\theta(1-m\sin^2(\theta))^{-1/2}{\rm ,}\qquad\mbox{and}\qquad m<1{\rm .}\nonumber
\end{eqnarray}
Using our defined quantities, from equation (7), we now write $\phi$ in terms of $u_p$, $u_a$, and $u_o$,
\begin{eqnarray}
\phi+c_1=\frac{2}{\sqrt{2Mu_o}}~F\left(\sin^{-1}\left(\sqrt{\frac{u-u_a}{u_p-u_a}}~\right),\frac{u_p-u_a}{u_o}\right)\label{eq9}
\end{eqnarray}
where 
\begin{eqnarray}
u_o=b+d=u_3-u_a=\frac{1}{2M}-(2u_a+u_p){\rm ,}\label{eq10}
\end{eqnarray}
and $c_1$ is a constant of integration to be determined later. Note the condition imposed upon the argument $(u_p-u_a)/u_o$ in 
equation (\ref{eq9}), namely $(u_p-u_a)/u_o <1$. It is the same condition 
given by equation (\ref{eq5}).  

Finally, the perihelion shift is simply given by
\begin{eqnarray}
\fl\Delta\phi&=2\left((\phi+c_1)|^{\phi_p}_{\phi_a}-\pi\right)=2\left(\phi_p-\phi_a-\pi\right)\nonumber\\
\fl&=2\left(\sqrt{\frac{2}{Mu_o}}K\left(\frac{u_p-u_a}{u_o}\right)-\pi\right){\rm ,}\quad\mbox{provided}\quad M<\frac{1}{2(2u_p+u_a)}{\rm ,}\label{eq11}
\end{eqnarray}
where $K(m)$ is a complete elliptic integral of the first kind, defined by
\begin{eqnarray}
\fl K(m)\equiv F\left(\frac{\pi}{2}|m\right)=\int^{\pi/2}_0{\rm d}\theta(1-m\sin^2(\theta))^{-1/2}{\rm ,}\qquad\mbox{and}\qquad m<1{\rm .}\nonumber
\end{eqnarray}

Standard notation, for the bound Kepler problem, employs the eccentricity $e$, the semi-latus rectum $p$, and the length of the semi-major axis $a$. The
relationships between these variables, and $u_p$ and $u_a$ are:  
\numparts
\begin{eqnarray}
u_p&=\frac{1+e}{p} \\
u_a&=\frac{1-e}{p} \\
\frac{2}{p}&=u_p+u_a \\
e&=\frac{u_p-u_a}{u_p+u_a}\\
2a&=\frac{1}{u_p}+\frac{1}{u_a}\\
\frac{1}{pa}&=u_p u_a\\
\frac{2e}{p}&=u_p-u_a\\
2ea&=\frac{1}{u_a}-\frac{1}{u_p}\\
p&=a(1-e^2){\rm .}\label{eq12} 
\end{eqnarray}
\endnumparts
Using these definitions, and $u_o$, given by
\begin{eqnarray}
u_o=\frac{1}{2M}-(2u_a+u_p)=\frac{1}{2M}-\frac{3-e}{p}{\rm ,}\label{eq13}
\end{eqnarray}
we can re-write equation (\ref{eq11}) using conventional notation. The condition, in equation (\ref{eq5}), 
is now written as
\begin{eqnarray}
\frac{M}{p}<\frac{1}{2(3+e)}\;{\rm,} \qquad\mbox{or}\qquad r_s<\frac{p}{3+e}\;{\rm.} \label{eq14}
\end{eqnarray}
At this point, we also write an expansion of $\Delta\phi$, for $M/p\ll1$,  
\begin{eqnarray}
\fl \Delta\phi&=2\left(\frac{2}{\sqrt{1-\frac{2M}{p}(3-e)}}\;K\left(\frac{4eM/p}{1-\frac{2M}{p}(3-e)}\right)-\pi\right)\label{eq15}\\
\fl&=\frac{6M\pi}{p}+\frac{3M^2\pi}{2p^2}(18+e^2)+\frac{45M^3\pi}{2p^3}(6+e^2)+...\;{\rm ,}\quad\mbox{for}\quad \frac{M}{p}<\frac{1}{2(3+e)}{\rm .}\label{eq16}
\end{eqnarray}

The first term, in equation (\ref{eq16}), agrees with the standard result, obtained using a perturbation technique to solve 
the non-linear ODE in equation (2a). It is expected that the exact result, given in equation (\ref{eq15}), will give
an insignificant correction when applied to the perihelion shift of Mercury, since $M_\odot/p\sim 10^{-8}\ll1$. A quick 
calculation for Mercury, using $p=185\;s$, $M_{\odot}=4.93 \times 10^{-6}\;s$, and $e=.210$, shows the small relative error:
\begin{eqnarray}
\left|\frac{\Delta\phi-\frac{6\pi M_{\odot}}{p}}{\Delta\phi}\right|=1.19 \times 10^{-7}{\rm .}\nonumber
\end{eqnarray}
The exact result will be advantageous for situations where $M/p$ is much larger. Since $e\in [0,1)$, this implies that 
$\frac{1}{2(3+e)} \in (\frac{1}{8},\frac{1}{6}]$. Therefore the largest value, implied by the condition in equation (\ref{eq14}), would
be $M/p\sim 10^{-1}$. The relative error is directly proportional to this parameter, specifically
\begin{eqnarray}
\frac{\Delta\phi-\frac{6\pi M}{p}}{\Delta\phi}=\frac{M}{4p}(18+e^2)+\frac{M^2}{16p^2}(36+24e^2-e^4)+...\;{\rm .}\nonumber
\end{eqnarray}  

\subsection{Bound Orbit Equation}

It is a simple matter to write down the orbit equation $u(\phi)$, simply invert equation (\ref{eq9}) to solve for $u$. The result is
\begin{eqnarray}
u(\phi)=u_a+(u_p-u_a)sn^2\left(\sqrt{\frac{Mu_o}{2}}(\phi+c_1),\frac{u_p-u_a}{u_o}\right)\label{eq17}
\end{eqnarray}
where $sn$, the Jacobi elliptic function, is defined by
\begin{eqnarray}
sn(v,m)\equiv sn(v|m)=sin(\varphi)\qquad\mbox{and}\qquad v=F(\varphi,m)\equiv F(\varphi|m)\;{\rm .}\nonumber
\end{eqnarray}
Common convention dictates that we choose $u(\phi=0)=u_p$. We therefore choose our integration constant, 
\begin{eqnarray}
\fl c_1=-\sqrt{\frac{2}{Mu_o}}K\left(\frac{u_p-u_a}{u_o}\right)\;{\rm ,}\qquad\mbox{such that}\qquad u(\phi=0)=u_p=\frac{1+e}{p}\label{eq18}
\end{eqnarray}
\begin{eqnarray}
\fl{\rm and}\qquad u\left(\phi=\sqrt{\frac{2}{Mu_o}}K\left(\frac{u_p-u_a}{u_o}\right)\right)=u_a=\frac{1-e}{p}\;{\rm .}\nonumber
\end{eqnarray}
Using equation (12), we now write the orbit $u(\phi)$ in terms of $p$ and $e$, along with its series expansion for $M/p\ll1$, as 
\begin{eqnarray}
\fl u(\phi)&=\frac{1+e\left(2~sn^2\left(\sqrt{1-\frac{2M}{p}(3-e)}\;\frac{\phi}{2}-K\left(\frac{4eM/p}
{1-\frac{2M}{p}(3-e)}\right),\frac{4eM/p}{1-\frac{2M}{p}(3-e)}\right)-1\right)}{p}\label{eq19}\\
\fl&=\frac{1+e\cos\phi}{p}+\frac{eM\sin\phi\;(3 \phi+e\sin\phi)}{p^2}+...\;{\rm ,}\quad\mbox{for}\quad \frac{M}{p}<\frac{1}{2(3+e)}{\rm .}\label{eq20}
\end{eqnarray}
The first term is, of course, the non-relativistic bound Kepler orbit for a simple $1/r^2$ radially inwards central force. The second term agrees with the standard
perturbation method result for this problem, the bound Kepler problem in GR.

\subsection{The $\frac{M}{p}<\frac{1}{2(3+e)}$ Condition}

As stated previously, we require $M/p<\frac{1}{2(3+e)}$ or $r_s<\frac{p}{3+e}$ to ensure the radial velocity is not imaginary. A practical exercise, at this
point, is to examine the invariants $E$ and $L$ to guarantee they satisfy this conditional statement. We also expect our invariants to approach their 
non-relativistic limit when $M/p\ll1$.

Equations (2b) and (2c) are solved simultaneously, giving
\begin{eqnarray}
\fl\left(\frac{E}{L}\right)^2=\frac{1}{M p}-\frac{4}{p^2}+\frac{4 M}{p^3}(1-e^2){\rm ,}\qquad\mbox{and}\qquad 
\left(\frac{m}{L}\right)^2=\frac{1}{M p}-\frac{3+e^2}{p^2}{\rm .}\nonumber
\end{eqnarray}
Isolating $E/m$ and $L/m$, we can re-write these equations as 
\begin{eqnarray}
\fl\left(\frac{E}{m}\right)^2=\frac{1-\frac{4 M}{p}+\frac{4 M^2}{p^2}(1-e^2)}{1-\frac{M}{p}(3+e^2)}{\rm ,}\qquad\mbox{and}\qquad 
\left(\frac{L}{m}\right)^2=\frac{M p}{1-\frac{M}{p}(3+e^2)}{\rm .}\label{eq21}
\end{eqnarray}
Equation (\ref{eq21}) implies that $(L/m)^2>0$ when $M/p<1/(3+e^2)$. The numerator, in the $(E/m)^2$ expression, is positive when 
$M/p>\frac{1}{2(1-e)}$ and $M/p<\frac{1}{2(1+e)}$, and negative when $\frac{1}{2(1+e)}<M/p<\frac{1}{2(1-e)}$, while the denominator
is positive when $M/p<1/(3+e^2)$, and negative when $M/p>1/(3+e^2)$. Thus it is readily apparent that both $(E/m)^2>0$ and $(L/m)^2>0$
when $M/p<1/(3+e^2)<\frac{1}{2(1+e)}$. This condition is less stringent than our original, i.e. $M/p<\frac{1}{2(3+e)}<1/(3+e^2)$, and is 
therefore satisfied. 

Finally, we would like to compare the limit of our invariants, for $M/p\ll1$, with the non-relativistic bound Kepler problem. For the non-relativistic
case, equation (\ref{eq3}) is represented by
\numparts
\begin{eqnarray}
\fl\left(\frac{du}{d\phi}\right)^2=(u-u_p)(u_a-u)=-u^2+\frac{2(E-U)}{m(L/m)^2}=-u^2-\frac{(1-e^2)}{p^2}+\frac{2u}{p}\;{\rm ,}\\
\fl\mbox{where}\qquad\qquad\quad\;\;\: U=-mMu \;{\rm ,}\\
\ \qquad E=-\frac{mM}{2a}=-\frac{mM(1-e^2)}{2p}\qquad\mbox{and}\\
\qquad \left(\frac{L}{m}\right)^2=Mp\;{\rm .}\label{eq22} 
\end{eqnarray}
\endnumparts
For the relativistic case, we can represent equation (\ref{eq3}), using the $\{p,e\}$ parameters, as
\begin{eqnarray}
\fl\left(\frac{du}{d\phi}\right)^2=2M(u-u_p)(u-u_a)(u-u_3)=-u^2+\frac{(\frac{E}{m})^2-1}{(L/m)^2}+\frac{2Mu}{(L/m)^2}+2Mu^3\nonumber\\
\fl\qquad\quad\;=-u^2-\frac{(1-\frac{4M}{p})(1-e^2)}{p^2}+\frac{2u(1-\frac{M}{p}(3+e^2))}{p}+2Mu^3\;{\rm ,}\label{eq23} 
\end{eqnarray}
where $(E/m)^2$ and $(L/m)^2$ are given in equation (\ref{eq21}). It's important to note that the parameter sets $\{E,L\}$, $\{u_p,u_a\}$, and 
$\{p,e\}$, i.e. the initial conditions, do not have the same functional dependence in equations (22) and (23), even though we are using the same variables. Although, we do expect them to be equal in the limit $M/p\ll1$, e.g. $\{E,L\}$, in equation (23), should satisfy 
$(L/m)^2\to Mp$ and $(E/m)^2-1\to -M(1-e^2)/p$ in this limit -- series expansions, using equation (21), confirm this assertion. Also, 
equation (23), if we neglect the $1/r^3$ potential term, reverts to equation (22) when $M/p\ll1$. 

\section{Conclusion}

Exact results for the perihelion shift $\Delta\phi$, and the orbital equation $r(\phi)$, for a bound test particle in the Schwarzschild metric were presented in a compact form that doesn't explicitly use the Weierstrass function. These simpler forms, using the complete elliptic integral of the first kind, and the Jacobi elliptic function, should prove useful in situations where the standard approximate forms, for $M/p\ll1$, are not applicable.

\end{document}